\documentclass{PoS}
\usepackage{myjournal}

\title{Combined upper limit on Standard Model Higgs boson production at CDF}

\ShortTitle{Combined upper limit on Standard Model Higgs boson production at CDF}

\author{\speaker{Adrian BUZATU}\thanks{On behalf of the CDF collaboration.}\\
        McGill University\\
        E-mail: \email{adrian.buzatu@mail.mcgill.ca}}

\abstract{The Higgs boson is the only elementary particle predicted by the Standard Model (SM) that has neither been confirmed nor refuted. The CDF collaboration has performed SM Higgs searches in many channels using $p\pbar$ collisions at a centre-of-mass energy $\sqrt{s}=1.96\tev$. We present the latest combined Higgs boson search at CDF. Since the previous year's combination, the sensitivity is increased through the addition of new channels, the improvement of existing channels and the addition of new data samples. We also use the latest parton distribution functions and $gg \rightarrow H$ theoretical cross sections when modelling the signal event yields. Using integrated luminosities of up to 8.2 $\invfb$, we observe a good agreement between data and the background prediction. Since we do not see a Higgs boson excess, we set 95\% CL upper limits on the Higgs boson cross section in the range between 100 and 200 $\gevcc$, with 5 $\gevcc$ increments. The observed (expected) limits for a 115 and a 165 $\gevcc$ Higgs boson are 1.55 (1.49) and 0.75 (0.79)~$\times$ SM, respectively. Since last year, the Higgs boson excluded range by CDF is extended to 156.5 - 173.7 and 100 - 104.5 $\gevcc$.}

\FullConference{The 2011 Europhysics Conference on High Energy Physics-HEP 2011,\\
		July 21-27, 2011\\
		Grenoble, Rh\^one-Alpes France}

\begin{document}

\section{Introduction}

\ \\The spontaneous symmetry breaking and the origin of mass for the charged fermions and the gauge bosons is explained in the Standard Model of elementary particles and their interactions (SM) by the Higgs mechanism. To test the validity of the mechanism, experimentalists need to test all its predictions. However, its sole testable prediction is the existence of a new elementary scalar particle, the Higgs boson. Confirming or refuting experimentally the Higgs boson has been the major task of experimental particle physics since the 1970s. The Higgs mechanism predicts the Higgs boson to have a non-zero invariant mass of an \emph{a priori} unknown value. For this reason one needs to search for the Higgs boson at all available mass values ($\mh$). Since the production and decay probabilities change as a function of mass, there are several analyses possible. The CDF collaboration has performed all such analyses and has combined them for maximal sensitivity. Higgs boson direct searches at the LEP experiments have set a 95\% CL lower limit of 114.4 $\gevcc$~\cite{LEPLimit}~\cite{LEP_EWK}. Indirect searches by precision electroweak fits have set a 185 $\gevcc$ upper limit. The remaining mass region is within the analysis capability of the CDF experiment using $p\pbar$ collision data at a centre-of-mass energy of $\sqrt{s}=1.96\tev$. Here we present the most recent CDF combination of direct searches in the range $100 < \mh < 200 \gevcc$.

\section{CDF Higgs Search Channels}

\ \\A total of 71 mutually exclusive final states are considered by the CDF, as explained in detail in~\cite{CDFCombination2011}~\cite{TevatronCombination2011}. The main analyses employed are $WH \rightarrow l\nu b\bbar$, $VH \rightarrow \MET b\bbar$, $ZH \rightarrow l^+l^-b\bbar$, $H \rightarrow \gamma \gamma$, $VH \rightarrow jj b\bbar$, $H \rightarrow W^+W^-$, $H \rightarrow \tau^+\tau^- jj$, $H \rightarrow ZZ \rightarrow llll$, two searches for $t\tbar H$ production, and a search for the associated production of $WH$ or $ZH$ with a tau lepton. In order to increase the sensitivity of each analysis, multivariate techniques such as artificial neural networks, boosted decision trees and matrix elements are used to compute the final discriminant. For all channels we normalize the Higgs boson prediction to the most recent higher order cross section calculations available. 

\begin{figure}[ht]
\begin{center}
\includegraphics[scale=0.3]{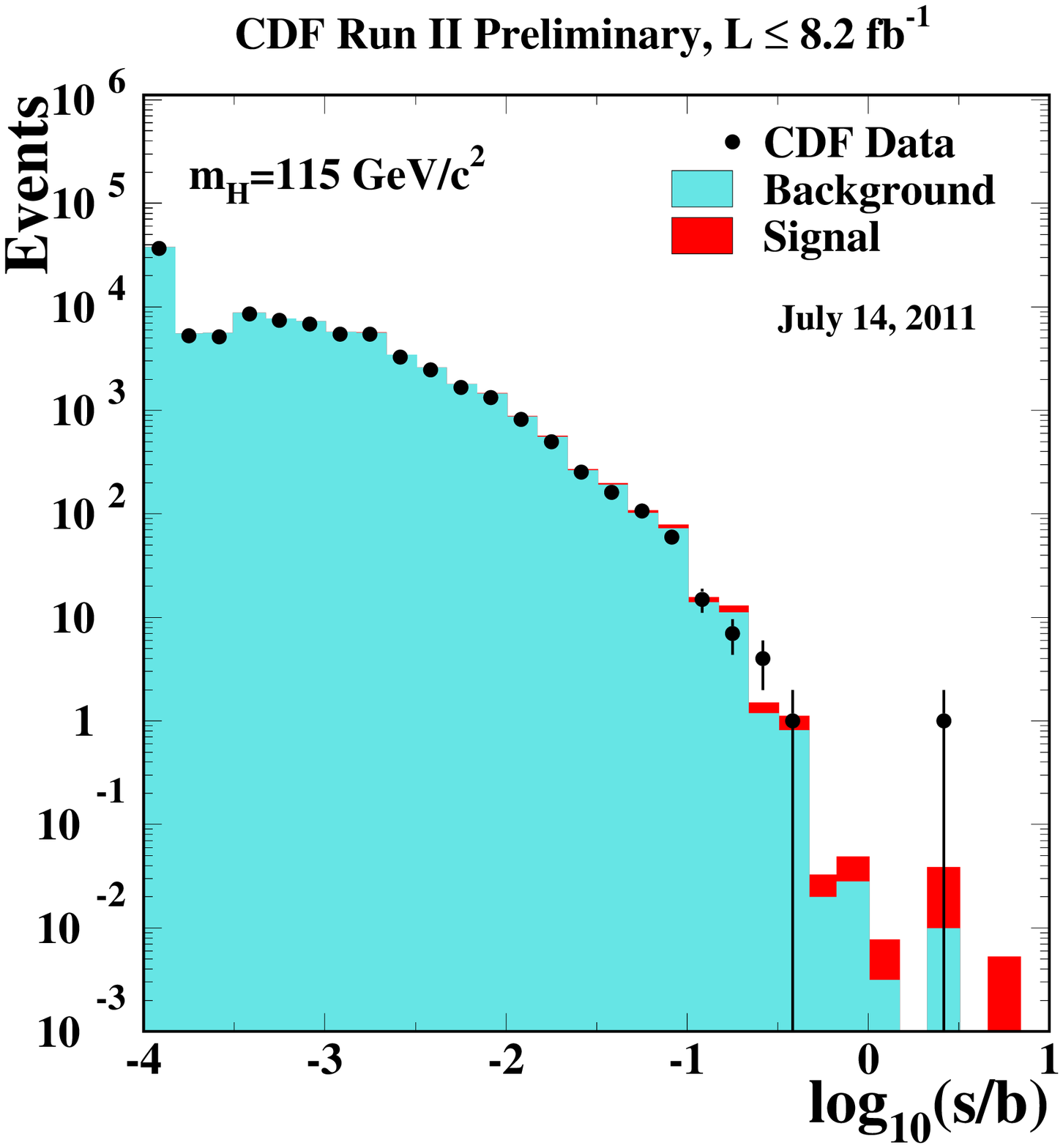}
\includegraphics[scale=0.3]{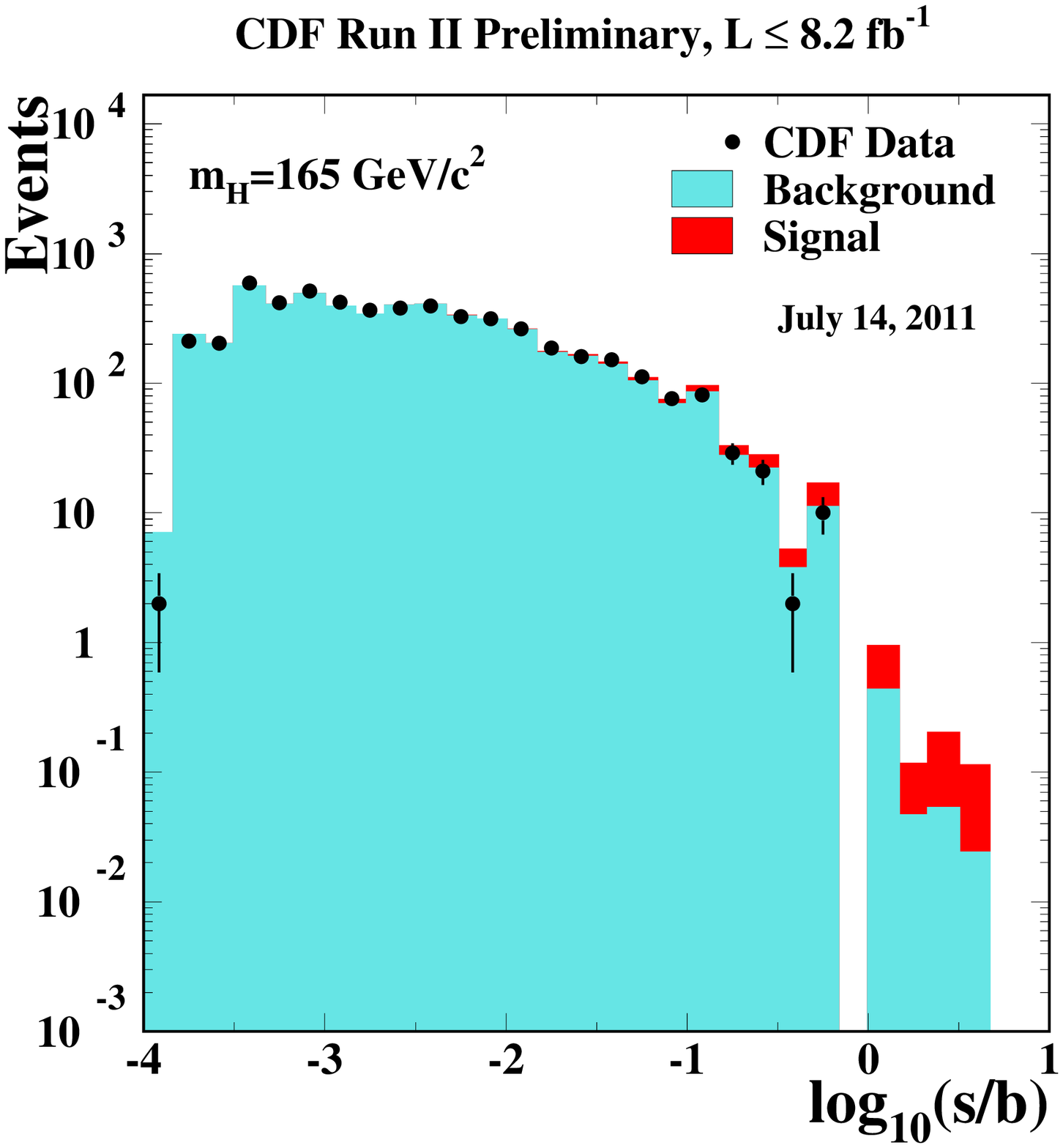}\\
\includegraphics[scale=0.3]{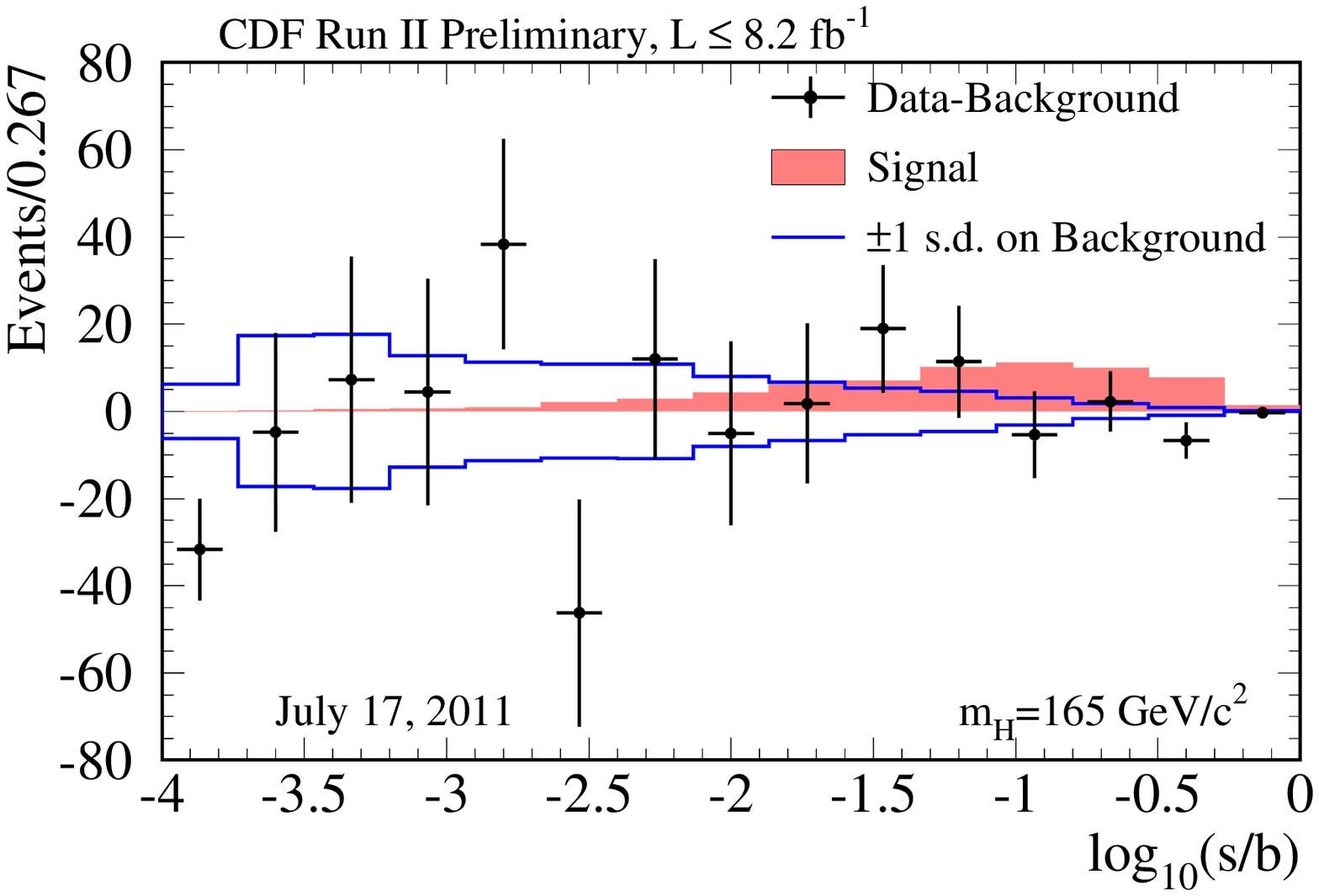}
\includegraphics[scale=0.3]{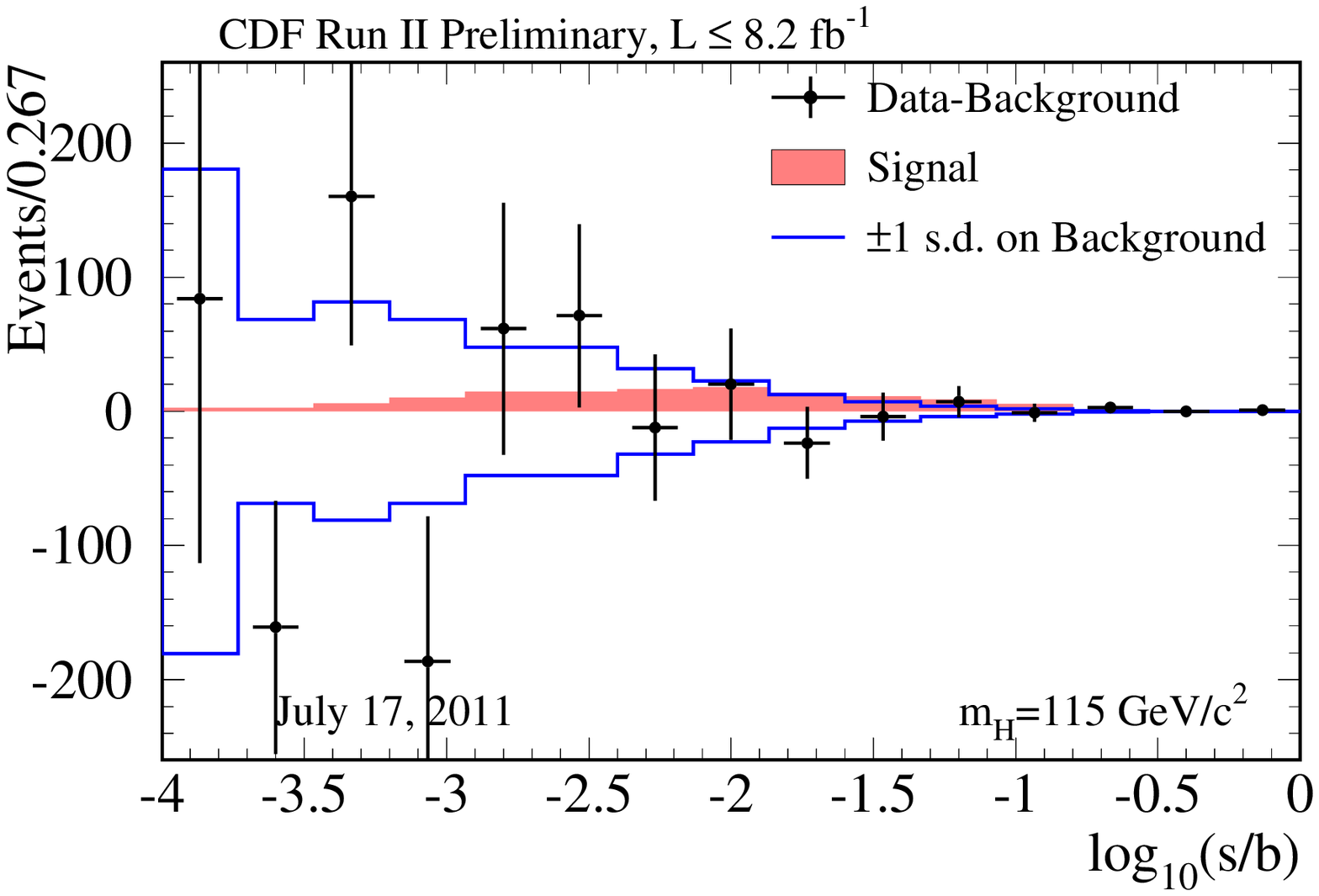}
\caption{Distribution of $log_{10}(s/b)$ (top) and background subtracted data distribution overlaid with Higgs signal (bottom) summed for all analyses used as input in the CDF combination for a $\mh=115\gevcc$ (left) and $\mh=165\gevcc$ (right).}
\label{figure:LogSB}
\end{center}
\end{figure}

\section{Combination Procedure}

\ \\The CDF combination employs a Bayesian statistical method with flat priors. All analyses use the full shape of their final discriminant in order to extract the most information possible from the distributions and therefore employ bin-by-bin Poisson statistics.

\ \\Two types of systematic uncertainties are used in the analyses. Rate systematics affect the overall discriminant normalizations. Some rate systematics are correlated (uncorrelated) among the analyses, such as the $b$-quark tagging, charged lepton and trigger efficiencies, integrated luminosity, background and signal cross sections (fake object identification, data-driven background modelling). Shape systematics affect the bin-by-bin normalizations and therefore the overall discriminant shapes, such as the jet energy scale.

\ \\We compare the data and background plus signal expectation by adding together all analysis final discriminants after having rebinned them as a function of the signal-to-background ratio ($s/b$), as seen in Figure~\ref{figure:LogSB} for $\mh=115\gevcc$ (top left) and $\mh=165\gevcc$ (top right). In such a plot, a signal would appear like an excess in data in the high $s/b$ bins. We also subtracted the background prediction from the data distribution and compared to the Higgs signal prediction, as seen in also Figure~\ref{figure:LogSB}, also as a function of $s/b$, both for $\mh=115\gevcc$ (bottom left) and $\mh=165\gevcc$ (bottom right).

\begin{figure}[ht]
\begin{center}
\includegraphics[scale=0.5]{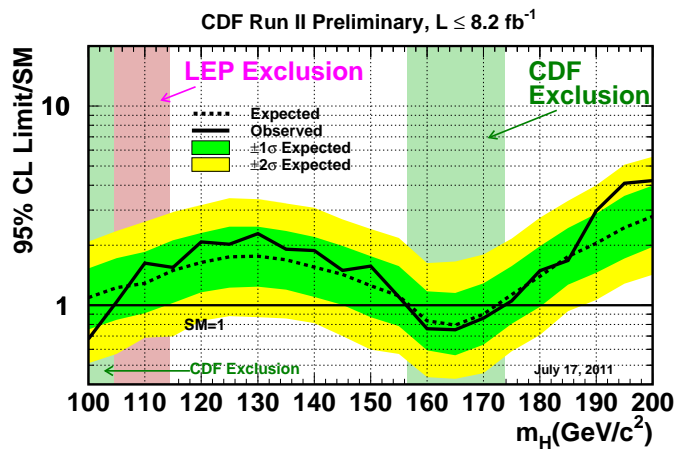}
\caption{Expected and observed Higgs boson cross-section 95\% upper limit as a function of the Higgs boson mass, between $100\gevcc$ and $200\gevcc$, with 5 $\gevcc$ increments. The horizontal line at 1 represents the SM prediction. The expected upper limits are represented by the dashed line. The yellow (green) band represents the 1 (2) standard deviation interval around the expected upper limit. The observed upper limits are represented by the solid line.}
\label{figure:Limit}
\end{center}
\end{figure}

\section{Results}

\ \\Since no significant excess of data over background is seen in bins with high $log_{10}(s/b)$, we conclude there is no evidence for a Higgs boson signal. We therefore proceed to compute observed and expected 95\% CL upper limits on the Higgs boson cross section divided by the Standard Model prediction in the range between 100 and 200 $\gevcc$, with 5 $\gevcc$ increments, as seen in Figure~\ref{figure:Limit}.

\section{Conclusions}

\ \\We have presented the most recent CDF direct Higgs boson search combination. Using integrated luminosities of up to 8.2 $\invfb$, we observe a good agreement between data and the background prediction. Since we do not see a Higgs boson excess, we set 95\% CL upper limits on the Higgs boson cross section in the range between 100 and 200 $\gevcc$, with 5 $\gevcc$ increments. The observed (expected) limits for a 115 and a 165 $\gevcc$ Higgs boson are 1.55 (1.49) and 0.75 (0.79) ~$\times$ SM, respectively. Since last year, the Higgs boson excluded range by CDF is extended to 156.5 - 173.7 and 100 - 104.5 $\gevcc$.

%\ \\The author would like to thank the conference organizers, the Tevatron accelerator team and the CDF collaboration.  

\ \\This work would not have been possible without the hard work of the Tevatron accelerator team and the CDF collaboration to produce and analyze the data.

\end{document}